# Quasiparticle Spin Resonance and Coherence in Superconducting Aluminium

C. H. L. Quay, Y. Chiffaudel, C. Strunk[1] and M. Aprili

Laboratoire de Physique des Solides (CNRS UMR 8502), Bâtiment 510, Université Paris-Sud, 91405 Orsay, France.
[1]Also at: Institute for Experimental and Applied Physics, University of Regensburg, 93040 Regensburg, Germany.

**Spin/magnetisation relaxation and coherence times, respectively $T_1$ and $T_2$, initially defined in the context of nuclear magnetic resonance (NMR), are general concepts applicable to a wide range of systems, including quantum bits [1-4]. At first glance, these ideas might seem to be irrelevant to conventional Bardeen-Cooper-Schrieffer (BCS) superconductors, as the BCS superconducting ground state is a condensate of Cooper pairs of electrons with opposite spins (in a singlet state) [5]. It has recently been demonstrated, however, that a non-equilibrium magnetisation can appear in the quasiparticle (i.e. excitation) population of a conventional superconductor, with relaxation times on the order of several nanoseconds [6-10]. This raises the question of the spin coherence time of quasiparticles in superconductors and whether this can be measured through resonance experiments analogous to NMR and electron spin resonance (ESR). We have performed such measurements in aluminium and find a quasiparticle spin coherence time of 95±20ps.**

If one thinks of spins as classical magnetic moments, $T_1$ is the time over which they align with an external magnetic field, while $T_2$ is the time over which Larmor-like precessions of the spins around the external field remain phase coherent [2, 11]. $T_1 \sim T_2$ for conduction electrons in most normal metals [3, 12, 13]. In a typical ESR experiment, electrons are immersed in an external homogenous static magnetic field, $H$. Microwave radiation creates a perturbative transverse magnetic field (perpendicular to the static field) of frequency $f_{RF}$. The power $P(H, f_{RF})$ absorbed by the spins from the microwave field is determined, usually by measuring the fraction of the incident microwaves that is *not* absorbed, i.e. either transmitted or reflected. When $H$ is tuned to its resonance value, $H_{res} = 2\pi f_{RF}/\gamma$ – with $\gamma$ the gyromagnetic ratio – the electron spins precess around $H$ and $P(H, f_{RF})$ is maximal. $P(H, f_{RF})$ is proportional to the imaginary part of the transverse magnetic susceptibility and to $[(H - H_{res})^2 + \frac{1}{(\gamma T_2)^2}]^{-1}$ in the case of a linearly polarised field [14]. Thus, $T_2 = 2/(\gamma \Delta H)$, where $\Delta H$ is the full-width at half-maximum of the power resonance as a function of $H$.

Our devices are thin-film superconducting (S) bars, with a native insulating (I) oxide layer, across which lie normal metal (N) electrodes [15], where S is aluminium, I is $Al_2O_3$ and N is thick aluminium with a critical magnetic field of ~50mT [16]. (In all the data shown here, this Al electrode is in the normal state.) A typical device, lying atop a Si/SiO$_2$ substrate, is shown in Figure 1. As in previous experiments, the NIS junctions have 'area resistances' of ~6·10$^{-6}$ Ω·cm$^2$ (corresponding to barrier transparencies of ~1·10$^{-5}$) and tunnelling is the main transport mechanism across the insulator. (See Supp. Info. of Ref. [8].) Measurements were performed at temperatures down to 60mK, in a dilution refrigerator. S has a superconducting gap of 205±10µV (based on conductance measurements between N1 and S), which corresponds to a critical temperature of 1.34±0.07K in the BCS theory.

A static magnetic field $H$ is applied in the plane of the device and parallel to the S bar (Figure 1a). The thickness of S ($d$ ~8.5nm) is well within the magnetic field penetration depth $\lambda$, which we expect to be ~315nm in our samples at 70mK. (See Supp. Info and Ref. [17] for details on this estimate.) The ratio of the orbital energy $\alpha = \frac{D(deH)^2}{6\hbar}$ to the Zeeman energy $E_z = \frac{1}{2}g\mu_B H$ is ~0.25 for the quasiparticles in S at the highest measured resonant magnetic field $H_{res}$ and is lower at lower fields. Therefore, the Zeeman energy is always dominant and we are in the 'paramagnetic limit' [18-20]. Here $D$ is the diffusion constant, $e$ the electron charge, $g$ the Landé *g*-factor, $\mu_B$ the Bohr magneton and $\hbar$ Planck's constant. (See Supp. Info. for details.)

A sinusoidal radio frequency signal of frequency $f_{RF}$ and rms amplitude $V_{RF}$ is applied across the length of the S bar (via a lossy, i.e. resistive coaxial cable). The resulting supercurrent flowing along the length of S serves primarily to produce the desired high-frequency magnetic field perpendicular to $H$; secondarily, it also breaks some Cooper pairs and thus increases the quasiparticle population. Microwave radiation due to the supercurrent thus impinges on the quasiparticle spins in S. Some of this radiation is absorbed by the quasiparticle spins, and the rest transmitted to and absorbed by the surrounding environment. The 'transmitted radiation' can appear as a voltage across a tunnel junction between S and N; this is the basis of our first detection scheme (DS1, Figure 1a). It can also be absorbed by the superconducting condensate, thus reducing the density of Cooper pairs and the current $I_S$ at which S becomes resistive, known as the switching current; this is the basis of our second detection scheme (DS2, Figure 1c). Both of our detection schemes for $P(H, f_{RF})$ are therefore entirely 'on-chip'.

In DS1 (Figure 1a), we apply a bias voltage $V_{DC}$ across an NIS junction and measure its differential conductance $G = dI/dV_{DC}$ using standard lock-in techniques. ($I$ is the current across the junction.) Figure 1b shows such traces as a function of $V_{RF}$, in which we see a flattening of the coherence peaks in a monotonic fashion. (This is similar to the effect of classical rectification [6].) Figure 1c shows a slice of Figure 1b at $V_{DC}$ = -288µV. $G$ across the junction can be seen to be an effective microwave power meter at the chosen operating point (red dot). We define $V_{RF}^0$ (for any given frequency) as the reference $V_{RF}$ (at the output of the generator) at which the effective voltage at the device is the same as that for $f_{RF}$ = 7.14GHz and $V_{RF}$ = 16.81mV. (See Supp. Info.)

In DS2 (Figure 1d), we measure the voltage-current characteristic of the S bar and record the switching current $I_S$. A current $I_{DC}$ is injected from one N electrode to another and the resulting voltage $V$ across the length of the bar is measured. Figure 1e shows the differential resistance $R = dV/dI_{DC}$ of the S bar as a function of $I_{DC}$ and of $V_{RF}$. The peaks in these traces correspond to $I_S$. $I_S$ can be seen to depend monotonically on $V_{RF}$ and is thus also a good measure of the latter.

Our on-chip detection provides improved sensitivity compared to earlier work on the spin resonance of conduction electrons in normal metals (CESR) [3, 21]. Indeed, based on calculations for CESR measurements on macroscopic samples, it was previously thought that CESR signals in Type I superconductors would be unmeasurably small [22]. This is no doubt why, while a considerable amount of work has been done on the CESR in normal metals since the 1950s, to our knowledge only one such measurement has been performed on a bulk BCS superconductor (Nb) in the vortex state, close to the critical field [23-25].

Having characterised our two microwave power meters, we now perform quasiparticle spin resonance (QSR) measurements using each of them in turn, and compare the results of both.

In the first set of measurements, using DS1, we operate the NIS junction detector – which is to say measure the differential conductance $G = dI/dV_{DC}$ across it – at a fixed $V_{DC}$ of -288μV and $V_{RF} = V_{RF}^0$. As can be seen in Figure 1c, at this operation point, small decreases in the absorbed power will result in a proportional increase in $G$. (It can also be seen that we remain in the linear regime in the measurements in Figure 2.)

Figure 2a shows $G(H)$ at several different $f_{RF}$. As expected, each trace shows a resonance, i.e. an increase in $G(H)$ due to the fact that more power is being absorbed by precessing quasiparticle spins and therefore less appearing across the SIS junction. We determine $H_{res}$ and $\Delta H$ by a Lorentzian with a linear-in-$H$ background signal to these data. The background comes from magnetic-field-induced orbital effects in the quasiparticle density of states. This measurement was repeated at different $f_{RF}$; $H_{res}$ as a function of $f_{RF}$ is shown in Figure 2b. A linear fit to the data gives a *g*-factor of 1.95±0.2, consistent with previous measurements of electrons in Al in the normal state [26, 27].

Note that $\Delta H$ may be larger than its intrinsic value if, for instance, the static field is inhomogeneous in the region of interest [2]. This would then lead to an underestimate of $T_2$. In our samples, however, $d/2 \ll \lambda$, as mentioned above. Thus, the magnetic field seen by the quasiparticles is homogeneous to ~ 1.5% in the superconductor, much smaller than the $\Delta H$ we measure, and our estimate of $T_2$ is unaffected by magnetic field inhomogeneity. This is confirmed by the fact that $\Delta H$ does not depend on $H_{res}$, as can be seen in Figure 2b. Field homogeneity has been a challenge for both ESR and NMR measurements performed on macroscopic Type II superconductors. In these, specimen dimensions greater than $\lambda$ mean that the field decays significantly within the specimen, and additional complications often arise from the presence of vortices.

In the second set of measurements we use DS2. As can be seen in Figure 1e, small decreases in the absorbed microwave power will result in a proportional increase in the switching current $I_S$. We first measure the differential resistance $R = dV/dI_{DC}$ of the S bar as a function of $I_{DC}$ and of $H$ at $f_{RF} = 6.05$GHz, $V_{RF} = 0.8V_{RF}^0$ (Figure 3a). We observe an increase in $I_S$ at $H = 0.17$T, which we identify as $H_{res}$ – at this field, the quasiparticle spins enter into resonant precession, thus absorbing more microwave power. Less power is then transferred to the superconducting condensate and $I_S$ increases.

In Figure 3b, we show $I_S$ as a function of $H$ at two different frequencies. ($I_S$ is the average of two hundred switching current values obtained from $V(I_{DC})$ measurements.) The expected resonance appears at both frequencies. To compare results from the two different detection schemes, we superimpose on these traces data from Figure 2a at the same $f_{RF}$. We see that both $H_{res}$ and $\Delta H$ are the same for both detection schemes. We also verified that $H_{res}$ and $\Delta H$ are independent of $V_{RF}$ (see Supp. Info. and black dots in Figure 2b).

We repeated the $V(I_{DC})$ measurements with a slightly different circuit: we applied current across the S bar while measuring the voltage difference between the N electrodes. We were not able to observe a resonance in this case. It would thus seem necessary to drive the quasiparticle population out of equilibrium to observe the quasiparticle spin resonance (QSR). This is expected: the more quasiparticles there are, the stronger the resonance signal should be.

As the switching current method is sensitive to a longer portion of the S bar compared to the tunnel junction method, the agreement between the two is confirmation that the magnetic field is quite homogenous along the entire length of the S bar between the two N electrodes. Thus, $T_2$ estimated form the resonance linewidth, while still a lower limit, should be reasonably close to the intrinsic value. That the linewidth measured by both methods is similar also means that $T_2$ is insensitive to the number of quasiparticles within a certain range. We estimate that the quasiparticle density is about two orders of magnitude higher in the supercurrent measurements (with injection across the tunnel junctions) than in the conductance measurements. (See Figure 1f, Supp. Info. and Ref. [28])

From the resonance linewidth (full width at half maximum), we obtain $T_2^S = 95\pm20$ps; this is fairly constant within the range of accessible fields (Figure 3b). Several remarks can be made about this result in comparison to $T_1^S$ (on the order of ns) in superconducting Al as well as $T_1^N \sim T_2^N$ in normal Al – 50ps in our previous work on 20nm-thick films at 4K [8] and of the same order of magnitude in other work [29-32]. (In comparing values for normal and superconducting states, it is thus important to consider samples of similar thickness: In normal Al, $T_1$ is lower in thin film samples compared to bulk samples due to surface scattering [8, 26, 27, 29-32].)

Thus, as the temperature is decreased and Al becomes superconducting, $T_1$ increases by two or three orders of magnitude whereas $T_2$ stays roughly the same.

$T_1^S \gg T_1^N$ means that inelastic spin relaxation processes are weaker in superconducting Al than in the normal state. To what extent this is due to the appearance of a superconducting state (as opposed to other temperature-dependent effects such as phonon-scattering) is unclear. If it is the case that $T_1^S \gg T_1^N$ is unrelated to superconductivity, we expect that this is due to impurity and band structure effects dominating phonon-related effects as the temperature decreases, especially as the former are particularly strong in Al as compared to other metals [3, 33-35].

In contrast, $T_2^S \sim T_2^N$ means that whereas phase decoherence processes are similar in both superconducting and normal Al, suggesting that spin decoherence (as opposed to spin relaxation) processes are unrelated to phonons and are also relatively unaffected by the superconducting pair potential.

That $T_1^N \sim T_2^N$ whereas $T_1^S \gg T_2^S$ tells us that spin decoherence and spin relaxation are strongly linked in normal Al [3] but not in superconducting Al. As the spin imbalances measured in normal Al are small compared to temperature [8], it could also be argued that there are no truly inelastic processes, hence $T_1^N \sim T_2^N$.

Further work on superconducting materials with different critical temperatures is needed to elucidate the influence of the superconducting state on $T_1$ and also on the $T_1/T_2$ ratio in both superconducting and normal states. Our techniques are compatible with both conventional and unconventional superconductors.

We note that our methods for measuring the coherence time can in principle be extended to other superconducting materials as long as they can be nanostructured.

**Methods**

We fabricate our samples with standard electron-beam lithography and angle evaporation techniques in an electron-beam evaporator with a base pressure of $5 \cdot 10^{-9}$ mbar. We first evaporate ~8.5nm of Al, which is then oxidised at $8 \cdot 10^{-2}$ mbar for 10' to produce a tunnel barrier, then 100nm of Al at an angle. We also evaporate 40nm of Co and 4.5nm of Al (as a capping layer) at another angle, but these electrodes were not used in this work. All transport measurements were done in a $^3$He-$^4$He dilution refrigerator with a base temperature of 60mK. Differential resistances were measured with standard loc-in techniques. The switching currents $I_S$ reported here are the mean values of 200 measurements.

**Acknowledgements**

We thank H. Hurdequint for helpful discussions on conduction electron spin resonance in normal metals, and M. Wiedeneder and R.W. Ogburn for helpful comments on the manuscript. This work was funded by an ANR Blanc grant (MASH) from the French Agence Nationale de Recherche. C.S. thanks the CNRS and the Université Paris-Sud for funding his sabbatical at the Laboratoire de Physique des Solides.

**Author Contributions**

C.Q.H.L., Y.C. and M.A. fabricated the samples and performed the measurements. All the authors contributed to the data analysis and the writing of the manuscript.

**Author Information**

The authors declare no competing financial interests. Correspondence and requests for materials should be addressed to C.Q.H.L. (charis.quay@u-psud.fr).

**Figure 1|Two on-chip microwave power detection schemes for superconducting (hybrid) devices. a, c,** Scanning electron micrograph of a typical device (scale bar = 1μm ) and schematic drawings of the measurement setups. In both cases a static magnetic field, $H$ is applied parallel to a superconducting bar (S, Al) and a sinusoidal signal of rms amplitude $V_{RF}$ and frequency $f_{RF}$ in the microwave range applied across the length of S (with a lossy coaxial cable in series), resulting in a high-frequency field perpendicular to $H$. To detect the spin precession of the quasiparticles in S, two on-chip detection methods are used. **a,** (Detection Scheme 1) A voltage $V_{DC}$ is applied between S and a normal electrode (N1, thick Al) with which it is in contact via an insulating tunnel barrier (I, $Al_2O_3$). The differential conductance $G = dI/dV_{DC}$ is measured, where $I$ is the current between N1 and S. **b,** $G$ as a function of $V_{DC}$ and nominal $V_{RF}$ (at the output of the generator and not accounting for attenuation in the lines). The red dot indicates the operation point of the detector for the data in Figure 2: $V_{RF} = V_{RF}^0$, $V_{DC}$ = -288μV. For any given frequency, we define $V_{RF}^0$ as the $V_{RF}$ (at the output of the generator) at which the effective voltage at the device is the same as that for $f_{RF} = 7.14$GHz and $V_{RF} = 16.81$mV. (See Supp. Info.) **c,** A slice of (b) at $V_{DC}$ = -288μV (blue dashed line in (b)), with the operation point indicated. **d,** (Detection Scheme 2) A current $I_{DC}$ is injected along the length of S. We measure either the voltage $V$ between the ends of the S bar or the differential resistance $R = dV/dI_{DC}$. We record in particular the switching current $I_S$ at which S first becomes resistive. **e,** $R$ as a function of $I_{DC}$ and nominal $V_{RF}$ (not accounting for attenuation in the lines). The switching current $I_S$ at which S become resistive appears here as a peak in $R$. $I_S$ can be seen to decrease monotonically with $V_{RF}$. The red dashed line indicates the operation point of the detector for the data in Figure 3: $V_{RF} = 0.8V_{RF}^0$. **f,** The blue trace is the first slice of (e) (blue dashed line in (e)), at $V_{RF} = 0.1$mV. The black trace is a two terminal measurement of the S bar, in the absence of microwaves, with a constant corresponding to the resistance of the lines subtracted. The difference in $I_S$ between the two indicates that the S bar is strongly out-of-equilibrium in our second (switching current) detection scheme. (See Supp. Info.)

**Figure 2|Spin resonance in conductance across tunnel junction. a,** NIS junction conductance $G$ as a function of $H$ at $V_{DC} = -288$μV and $V_{RF} = V_{RF}^0$ for different $f_{RF}$. The black vertical line indicates the critical field of N. $H_{res}$ and $\Delta H$ are obtained for each $f_{RF}$ by fitting a Lorentzian with a linear background. The fit for $f_{RF} = 10.56$GHz is shown (thin red line) and $H_{res}$ indicated with a red vertical line. **b,** $H_{res}$ and $\Delta H$ the resonance linewidth (full-width at half-maximum) as a function of $f_{RF}$. A linear fit to $H_{res}(f_{RF})$ data gives a Landé g-factor of 1.95±0.2. The black dots indicate values obtained at different powers or with the second detection scheme. (See Supp. Info.)

**Figure 3|Spin resonance in supercurrent, comparison of detection schemes. a,** Differential resistance $R$ of the S bar as a function of $H$ and $I_{DC}$ with $f_{RF} = 6.05$GHz, $V_{RF} = 0.8V_{RF}^0$. At $H_{res} \sim 0.17$T, the resonant field, the switching current $I_S$ can be seen to increase, indicating that less microwave power is being transmitted to the superconducting condensate

(see Supp. Info.) as more power is absorbed by the quasiparticles in S. **b,** Switching current $I_S$ as a function of static magnetic field $H$ for two different $f_{RF}$. (Here a slight change was made to the measurement circuit: With reference to Figure 1d, the current is applied between N2 and F instead of between N1 and N2, hence the slightly higher $I_S$: the current injection electrodes are closer together.) Superimposed on these traces are the conductance traces from Figure 2a at the same fields. $H_{res}$ and $\Delta H$ can be seen to be similar for both measurement methods. The bold red trace has been offset downwards by 19nA. **c,** The switching currents in (b) averages of ~200 measurements, with a standard deviation of about 3nA. Here we show a histogram of 250 switching currents corresponding ot the first point in the bold red trace in (b). Current was driven long the length of the S bar and the voltage measured between N1 and N2. (Voltage and current leads are thus switched with respect to (b).)

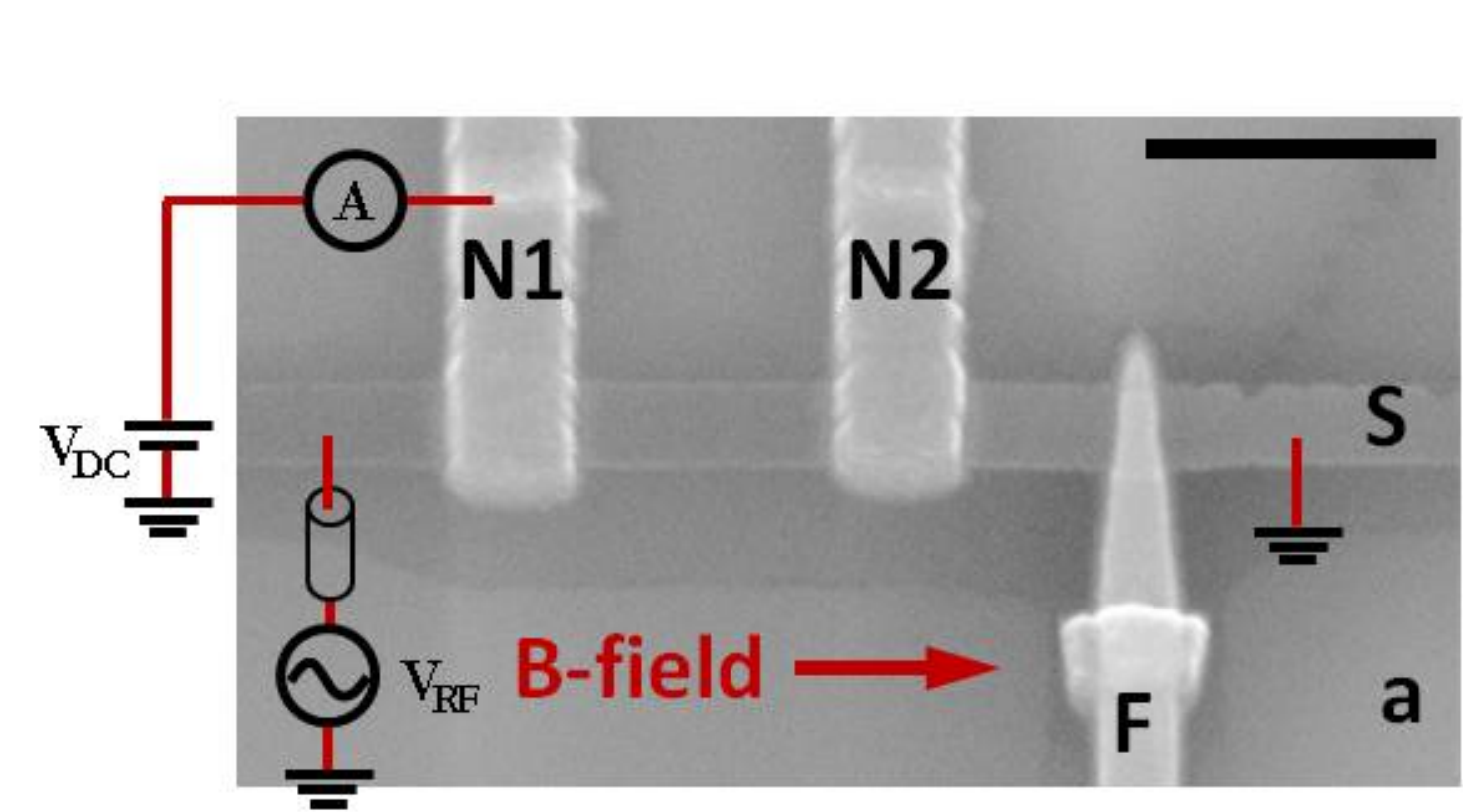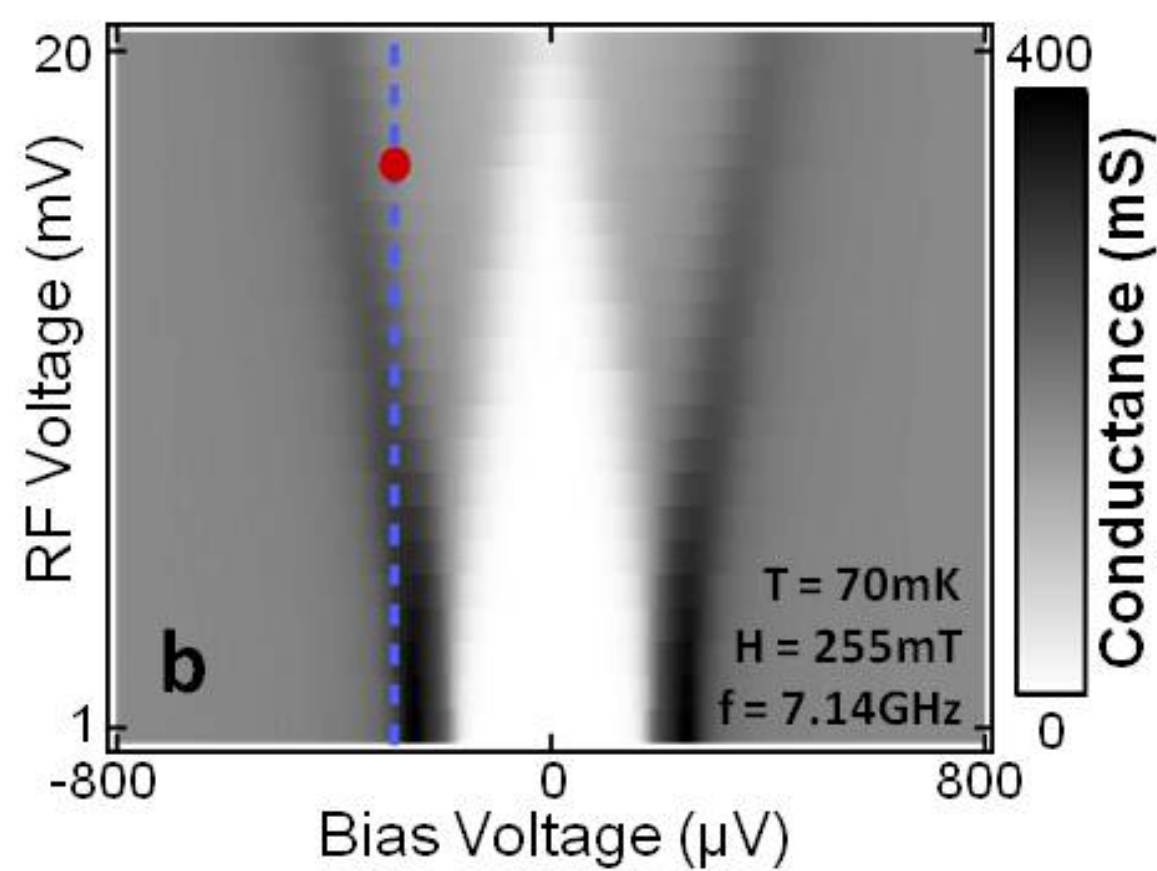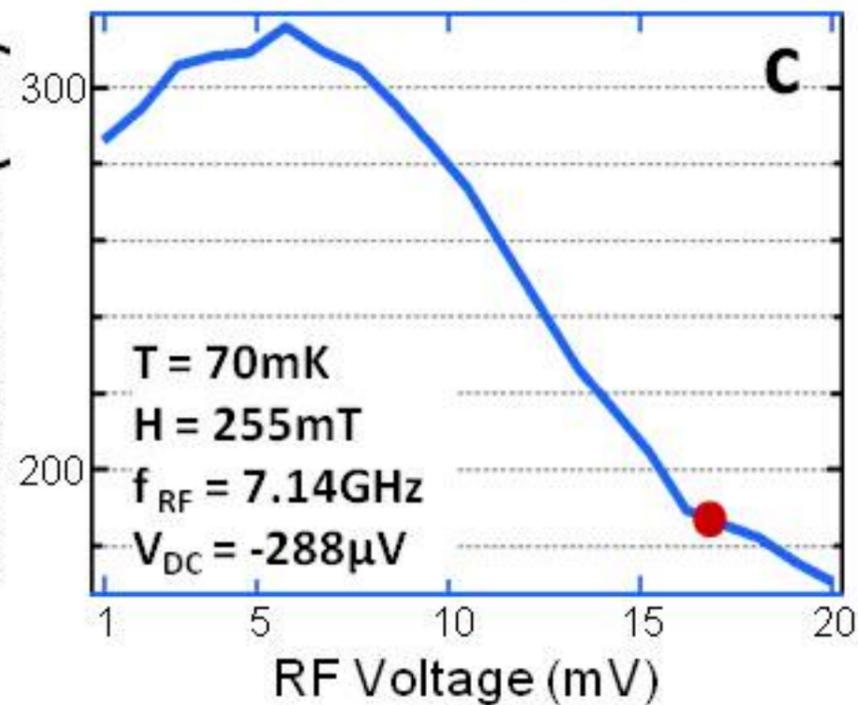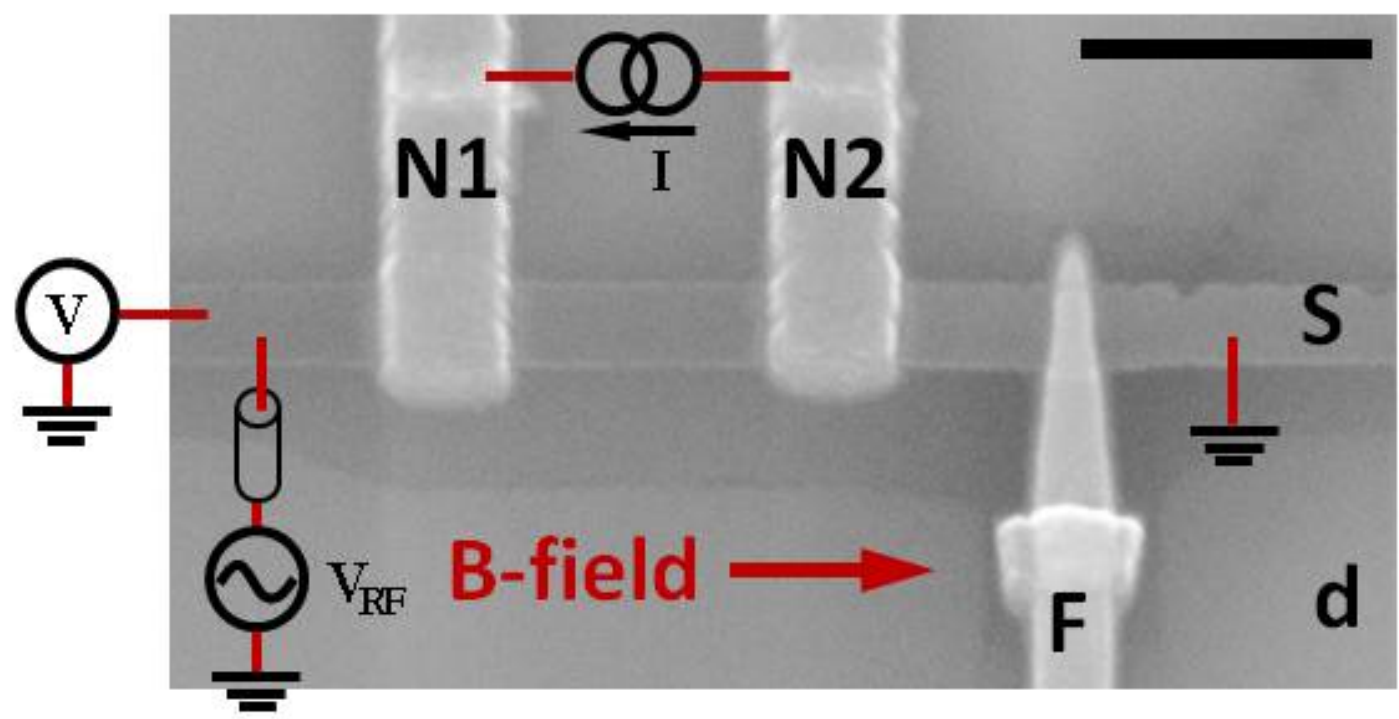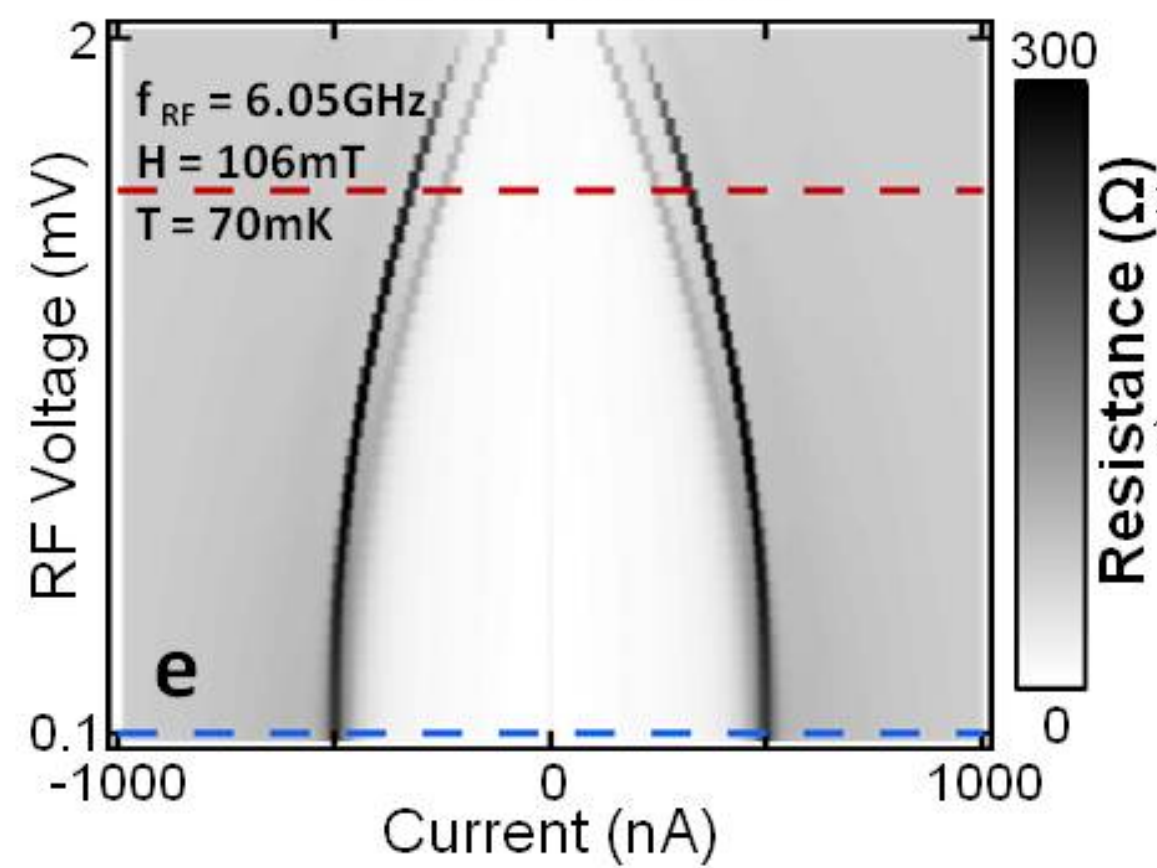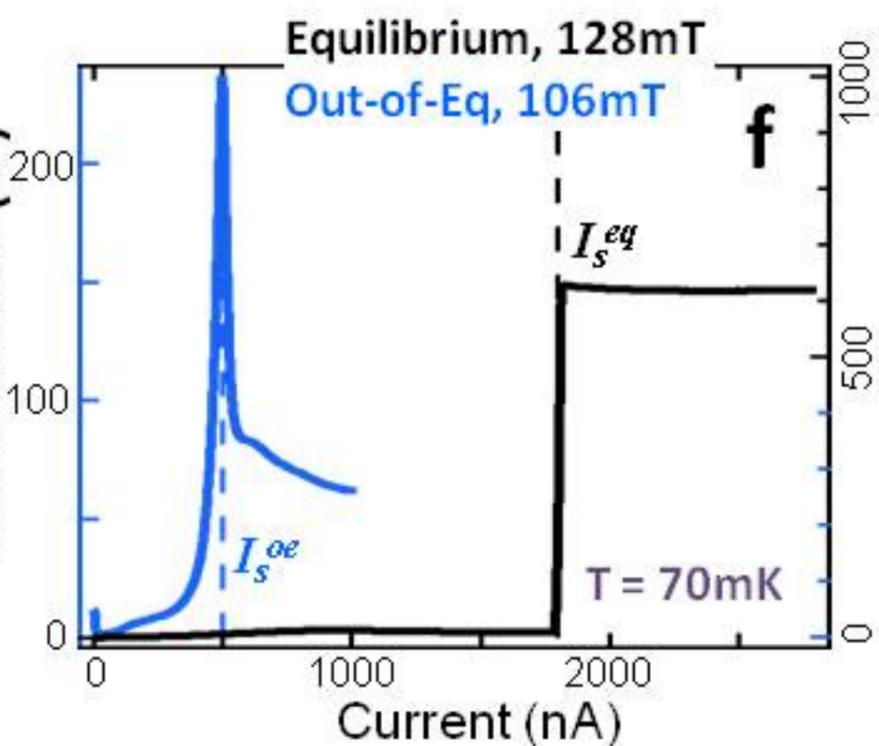

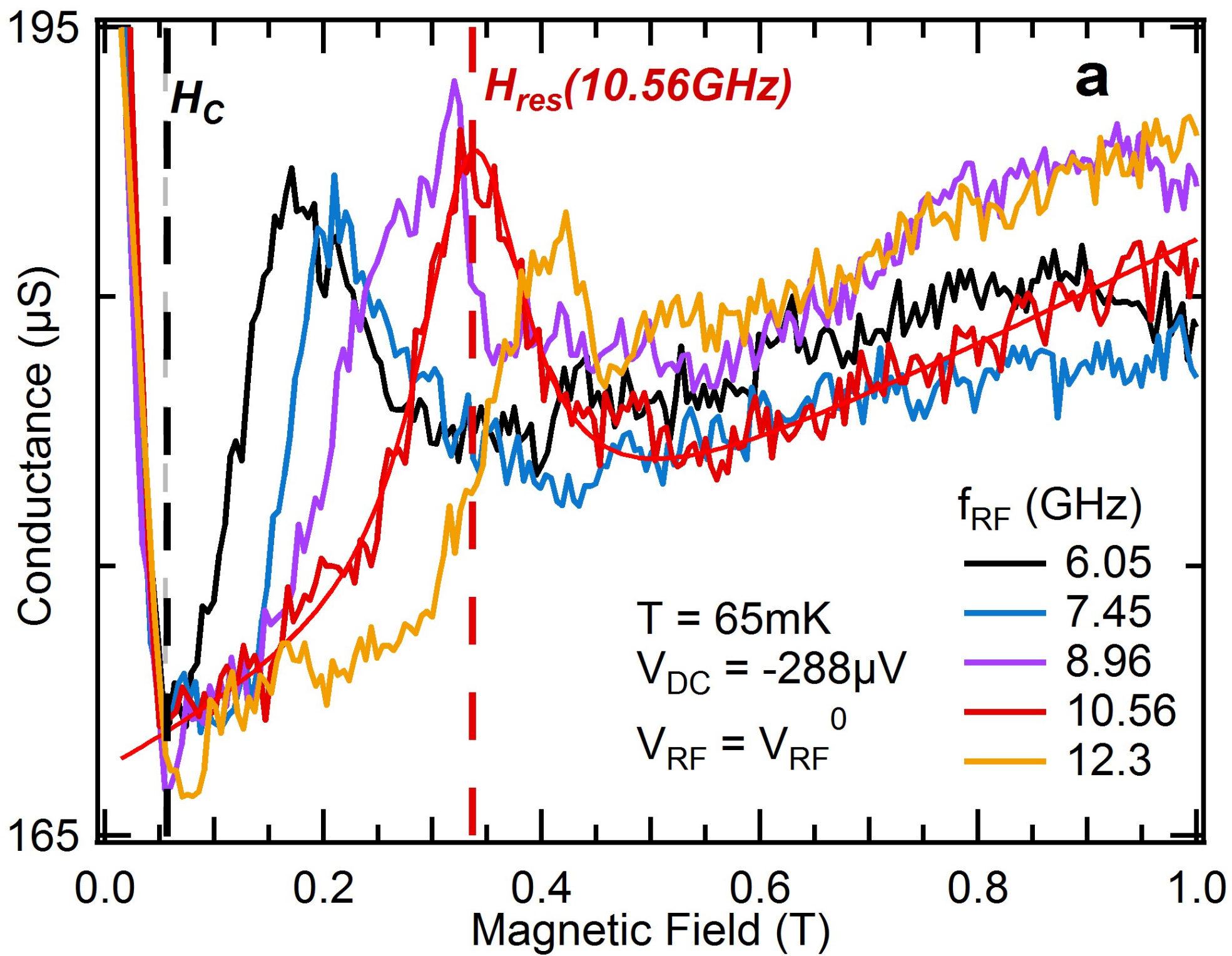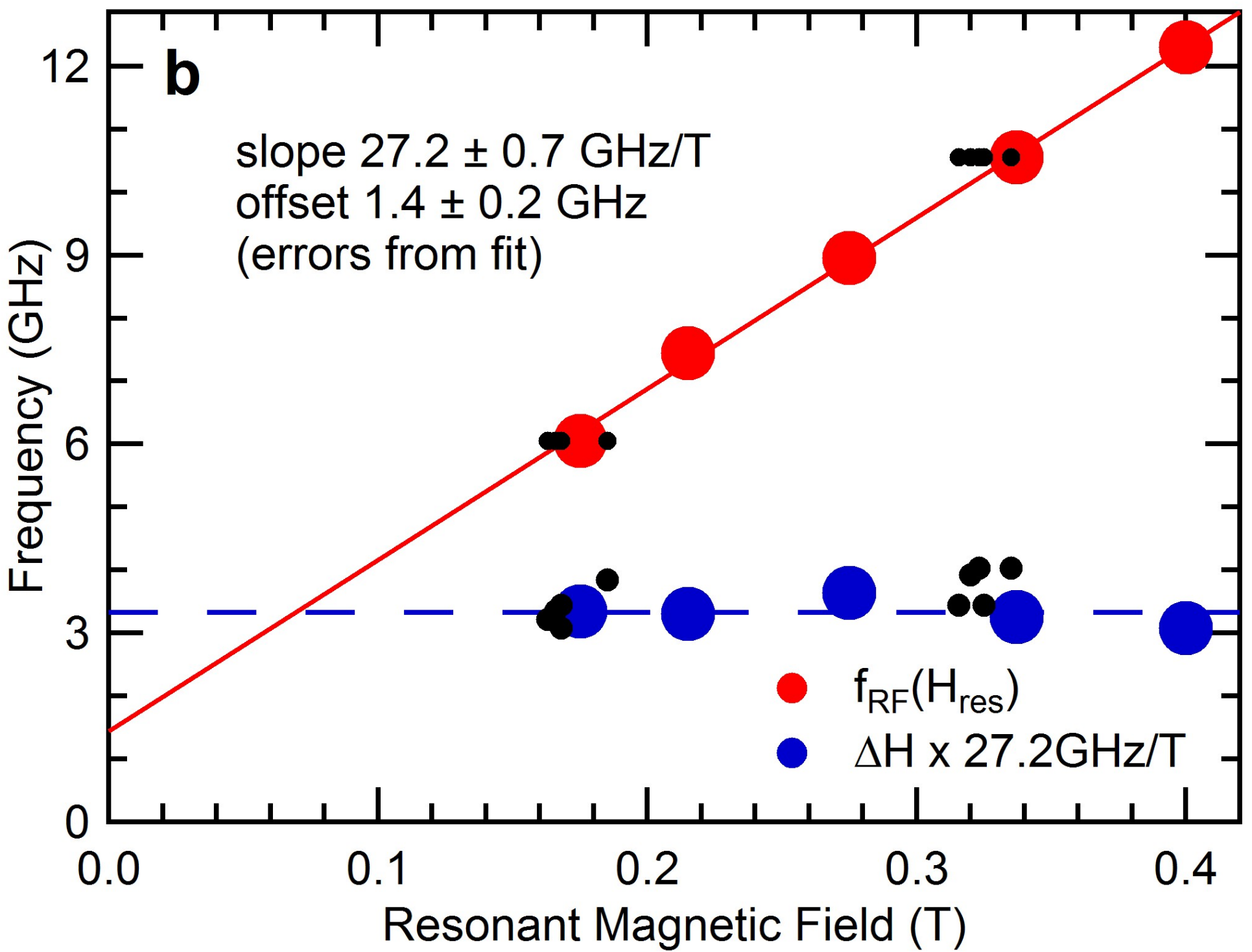

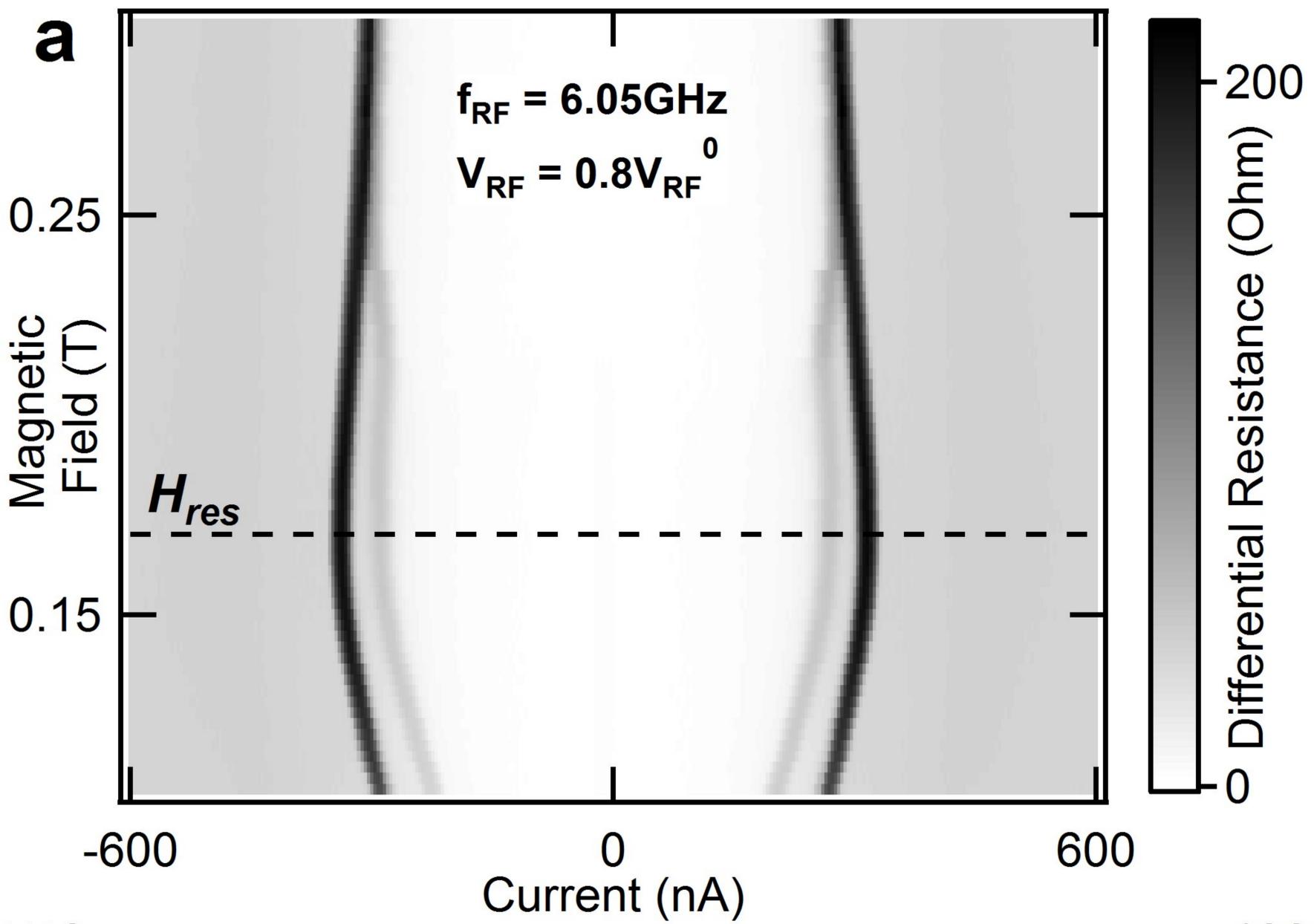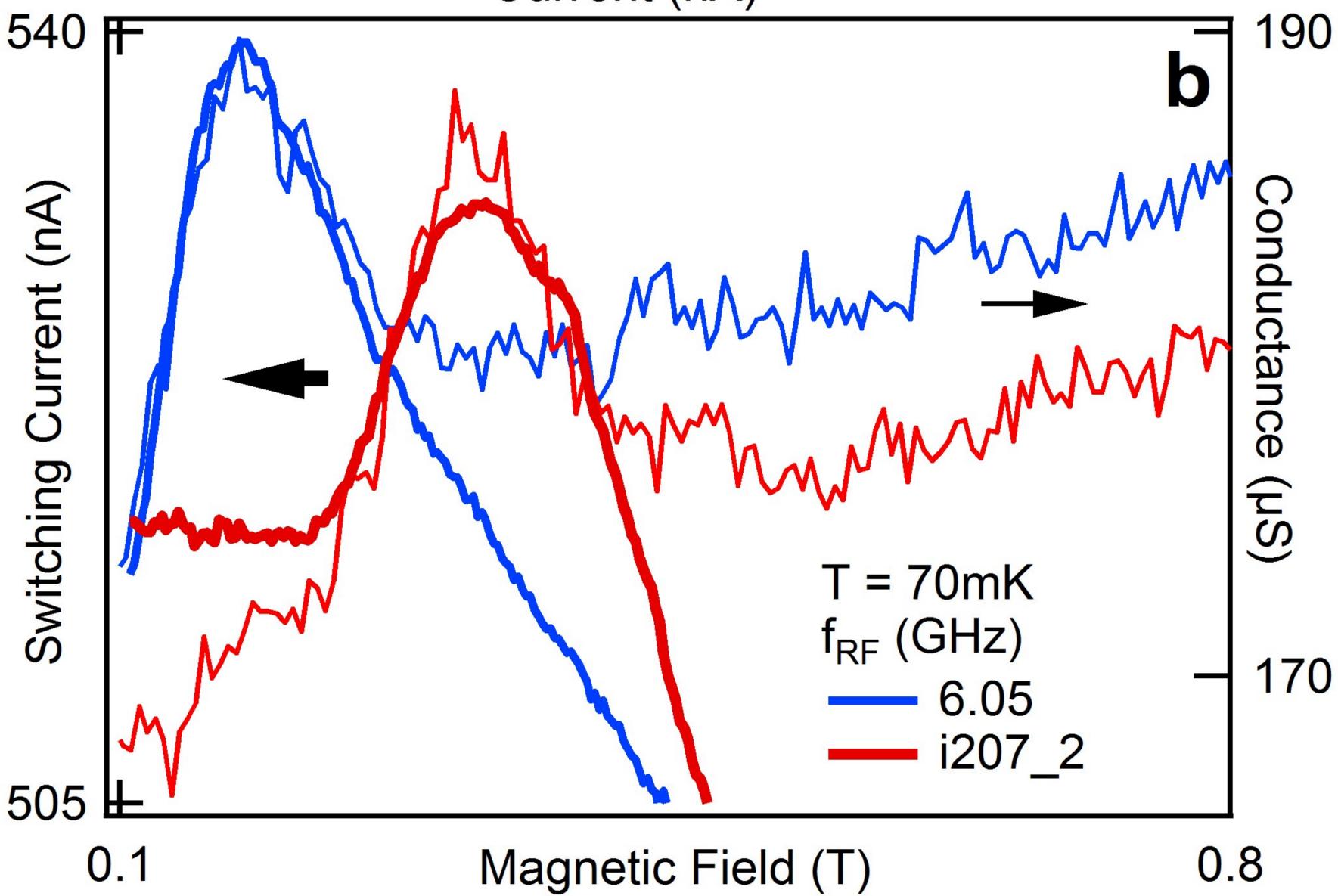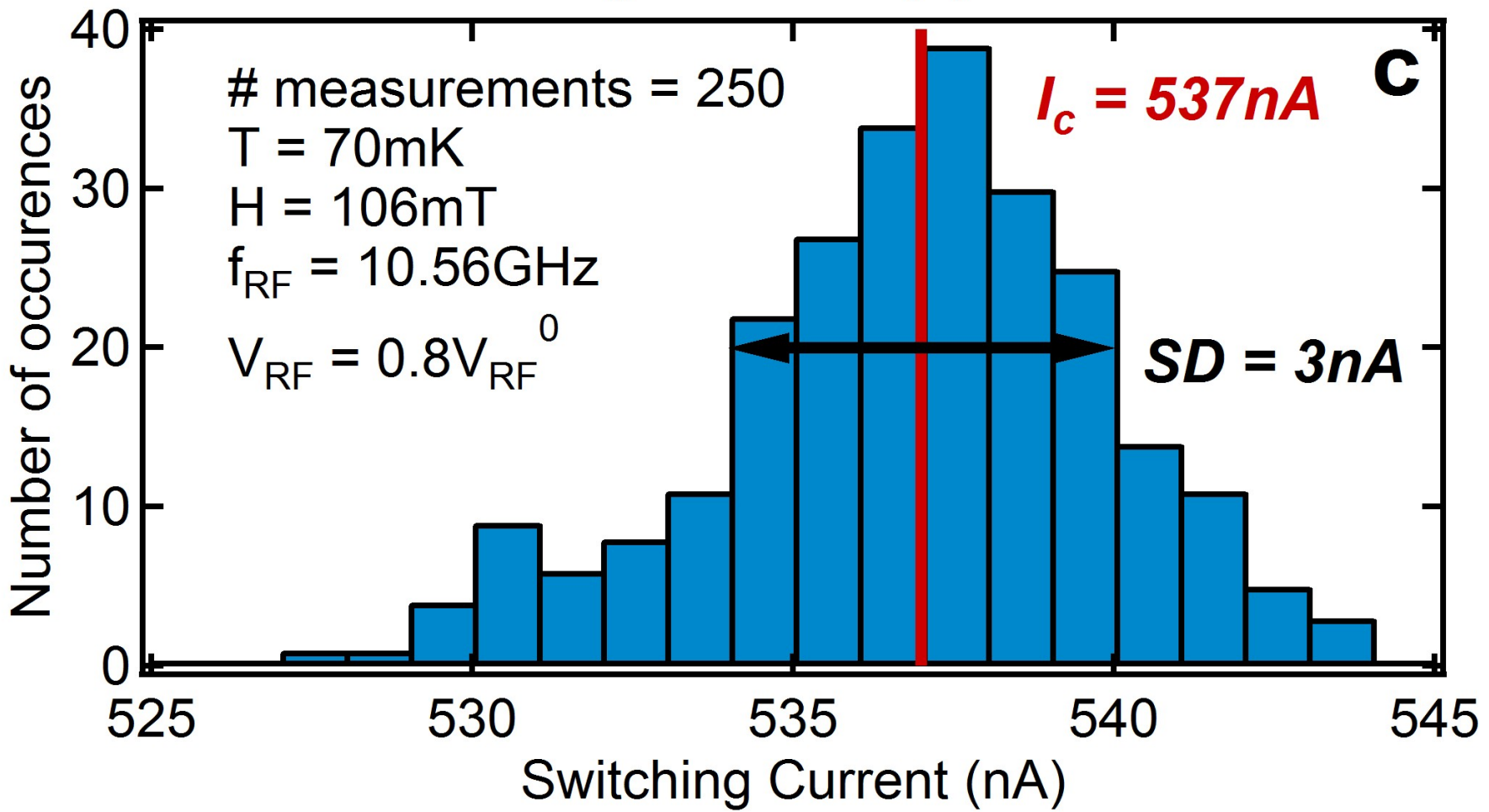

# Supplementary Information for 'Quasiparticle Spin Resonance and Coherence in Superconducting Aluminium'

C. H. L. Quay, Y. Chiffaudel, C. Strunk[1] and M. Aprili

Laboratoire de Physique des Solides (CNRS UMR 8502), Bâtiment 510, Université Paris-Sud, 91405 Orsay, France.
[1]Also at: Institute for Experimental and Applied Physics, University of Regensburg, 93040 Regensburg, Germany.

## A. Conductance-Voltage Traces in the Absence of Microwaves

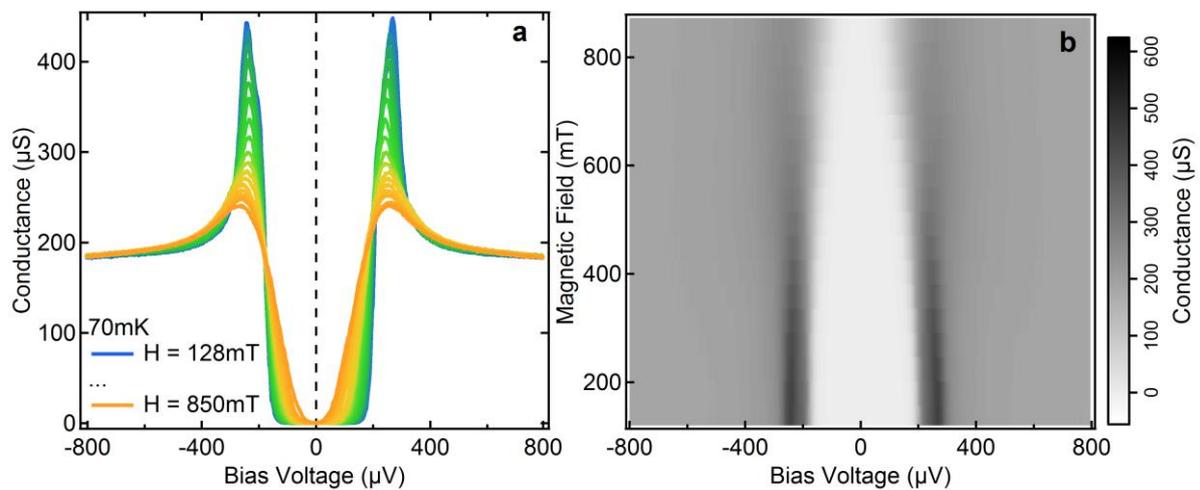

**Figure 1|** Conductance as a function of bias voltage and magnetic field across the junction N1-S in the absence of microwaves. The same data are plotted in two different ways.

Figure 1 shows the conductance of the junction N1-S, used as the detector for Figure 2 of the main text, as a function of bias voltage and magnetic field, above the critical field of N1. (N1 is normal for all the data shown in and relevant to the points made in the main manuscript.) The conductance spectra can be seen to vary smoothly as a function of magnetic field.

## B. Estimates of $D, \lambda$ and $\alpha$ in the superconducting Al film

The conductivity $\sigma$ of a diffusive metal is related to the diffusion constant $D$ by $\sigma = e^2 ND$, where $e$ is the charge of the electron and $N$ the density of states [1]. We obtain $\sigma = 8.2 \times 10^6$S/m from the blue trace in Figure 1f of the main text, taking the relevant volume to be that of the S bar between the electrodes N1 and N2 from centre to centre. Using $N = 2.4 \times 10^{22}$ states per eV per cm³ for aluminium, we obtain $D = 2 \times 10^{-3}$m·s$^{-2}$.

The penetration depth of the magnetic field into the superconductor is $\lambda = \sqrt{\frac{\hbar}{\mu_0 \pi \sigma \Delta}} \sim$ 315nm [2]. Here $\mu_0$ is the vacuum permittivity, $\hbar$ Planck's constant and $\Delta$ the zero-temperature superconducting gap. $\lambda$ is much greater than the thickness of our Al film, $d \sim$8.5nm.

The orbital energy of electrons in diffusive thin films with a magnetic field $H$ applied in the plane of the film is $\alpha = \frac{D(deH)^2}{6\hbar}$ whereas the Zeeman energy is $E_z = \frac{1}{2}g\mu_B H$, with $g$ the Landé $g$-factor and $\mu_B$ the Bohr magneton [3, 4]. (Note that Ref. [3] uses cgs units and Ref. [4] is missing a factor of $\hbar$. The expressions given here are correct and in SI units.) At the highest resonant field (~0.4T) measured in this work, we have $\alpha/E_z$~0.25. We are thus always in the 'paramagnetic limit', where the Zeeman effect dominates over orbital effects.

### C. Choice of Frequencies

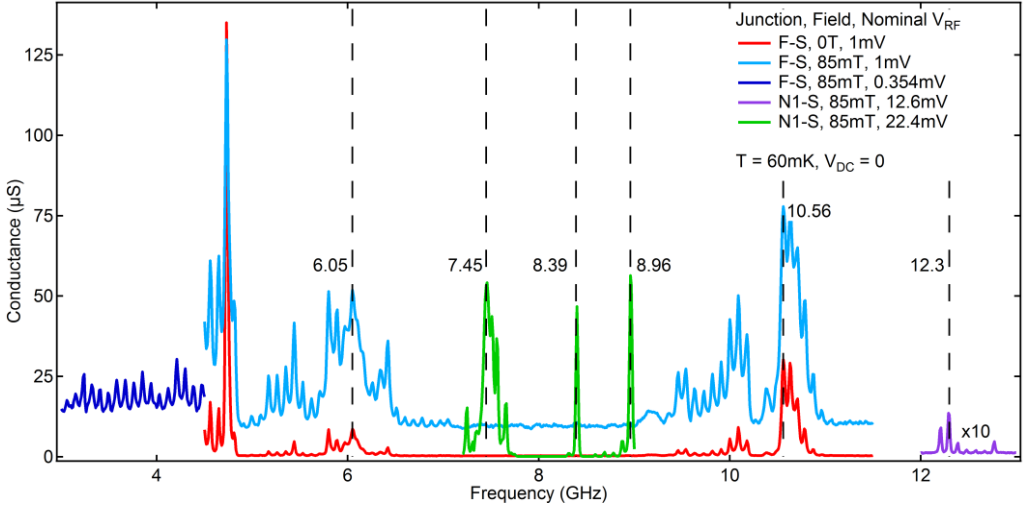

**Figure 2|** Zero-bias conductance of two different junctions as a function of frequency for several different nominal microwave voltages (at the output of the generator).

As in our previous work [5], Figures 1b and 1e of the main text are reproducible at any frequency modulo a constant shift in the $V_{RF}$ axis, with $V_{RF}$ being nominal microwave voltage (i.e. the voltage at the output of the generator). This constant shift is due to the frequency-dependent attenuation of our microwave line (greater attenuation at higher frequencies) as well as resonances in the line. The conductance of a junction at zero bias is thus a measure of the microwave power arriving at the junction/device.

As noted in our main text, we define $V_{RF}^0$ (for any given frequency) as the nominal $V_{RF}$ at which the effective voltage at the device is the same as that for $f_{RF} = 7.14$GHz and $V_{RF} = 16.81$mV.

To select the frequencies at to search for the quasiparticle spin resonance we measure the conductance of a junction as a function of frequency at various $V_{RF}$ and at $V_{DC} = 0$. (Figure 2) As can be seen in Figure 1b of the main text, the conductance at $V_{DC} = 0$ has a monotonic dependence on the effective $V_{RF}$ at the device and can be taken as an indication of the latter. The effect of the frequency-dependent transmission of our microwave line is quite apparent in these data.

For the measurements shown in the main text, we selected $f_{RF}$ at which the conductance is at a locally maximal, corresponding also to local maxima in the real microwave voltage at the device as a function of $f_{RF}$. We do this to avoid any experimental missteps accidentally delivering more power to the device than required, thus possibly blowing it up. In addition, this avoids unnecessary dissipation of energy in the microwave line, which if excessive could lead to a rise in the base temperature of the dilution refrigerator. (We did not notice this in our measurements.)

## D.   Quasiparticle Spin Resonance, Dependence on Microwave Amplitude (Both Detection Schemes) and on (Detection Scheme 1)

As explicated in the main text, the 'operating point' of Detection Scheme 1 (DS1) is defined by the chosen values of $V_{RF}$ and $V_{DC}$ whereas the operating point of Detection Scheme 2 (DS2) is defined by the chosen value of $V_{RF}$. We show here that our results for resonant field $H_{res}$ and the resonance linewidth $\Delta H$ are robust against the choice of operating point in both detection schemes.

We first show that our results from both detections schemes are independent of $V_{RF}$. Figure 3 shows the resonances from Figure 3b of the main text, together with the same measurements taken at different $V_{RF}$. Traces taken with DS1 are shown in Figure 3a while those taken with DS2 are shown in Figure 3a. Visually, $H_{res}$ and $\Delta H$ are not significantly affected by the microwave voltage. The values for $H_{res}$ and $\Delta H$ that we obtain from the fits shown here are plotted in Figure 3c of the main text (black dots).

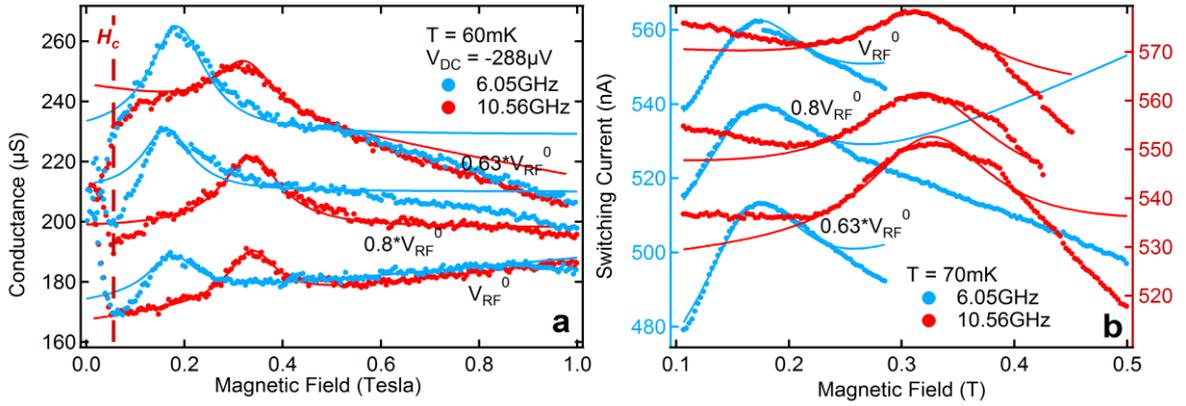

**Figure 3**| Quasiparticle spin resonances at different microwave voltages, measured with. **a,** Detection Scheme 1. **b,** Detection Scheme 2. For DS2, the measurement circuit is the same as in Figure 3b. Solid lines are fits of Lorentzians with a linear background.

Next, we show that our results from DS1 are independent of the choice of $V_{DC}$. In Figure 4a, we show measurement of conductance as a function of applied magnetic field at $V_{RF} = 0.8V_{RF}^0$, $V_{DC}$= -288μV (black trace in 4a, black dot in 4b) together with the same measurement taken at $V_{RF} = 0.8V_{RF}^0$, $V_{DC}$= -100μV (blue trace in 4a, blue dot in 4b). For the black trace we have $H_{res}$ = 340mT±5mT and $\Delta H$ = 148mT±25mT while for the blue trace we have $H_{res}$ = 340mT±5mT and $\Delta H$ = 154mT±25mT.

The resonance appears as a peak in the blue trace and a dip in the black trace. This can in fact be understood by looking at Figure 4b: As explicated in the main text, at the resonance, some of the microwave radiation is absorbed by the quasiparticle spins and so less is transmitted to the detectors. We can see from Figure 4b that at $V_{DC}$= -288μV a smaller effective $V_{RF}$ gives a higher conductance, whereas at $V_{DC}$= -100μV a smaller effective $V_{RF}$ gives a lower conductance, hence the different in the sign of the resonance in the two traces. To optimise sensitivity for this detection scheme, $V_{DC}$ and $V_{RF}$ should be chosen so that $dG/dV_{RF}$ is maximal.

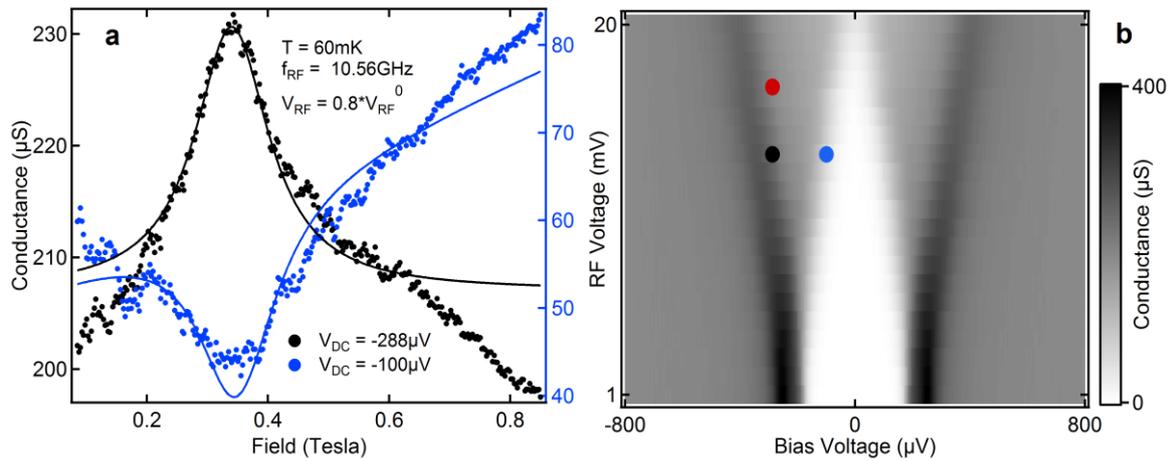

**Figure 4| a,** The quasiparticle spin resonance measured with Detection Scheme 1 at the operating points shown in (b). **b,** The colour-coded operating points for the blue and black traces shown in (a). The red dot is the operating point for the data shown in Figure 2a of the main text.

### E. Estimate of the Number of Quasiparticles

The switching current $I_s$ of the Al bar in the absence of microwaves and of quasiparticle injection is ~1800μA (Figure 1f of the main text): we remind the reader that, for the blue trace, current is injected along the length of the S bar. Detection Scheme 1 should be close to this 'equilibrium' situation as the voltages applied across the NIS junctions are of the order of the superconducting gap Δ (at zero temperature).

In contrast, in Detection Scheme 2 (Figure 1f and Figure 3 of the main text), current is injected into the S bar across a tunnel junction and 'removed' via another such junction, e.g. as shown in Figure 1d of the main text. Here, the voltages across the NIS junctions (which typically have resistances of ~5kΩ) at the point where the S bar becomes normal are several mV and we expect the quasiparticles in the S bar to be driven strongly out-of-equilibrium by the injected current. Typical $I_s$ measured are around 500-600nA.

If we assume that the non-equilibrium quasiparticle population can be described by an effective temperature $T_{eff}$, and that $I_s$ scales with $T_{eff}$ in the same way that Δ does, then in DS2 $T_{eff} > T_c/2$, with $T_c$ being the critical temperature. Based on Figure 4 of Ref. [6], we can then say that the quasiparticle density in DS2 is at least two orders of magnitude higher than in DS1.